\documentclass{webofc}
\usepackage[varg]{txfonts}   
\woctitle{Physics at the Magnetospheric Boundary, Geneva, Switzerland (25-28 June, 2013)}
\def\beq#1{\begin{equation}\label{#1}}
\def\eeq{\end{equation}}
\def\beqa#1{\begin{eqnarray}\label{#1}}
\def\eeqa{\end{eqnarray}}

\def\Eq#1{(\ref{#1})}

\def\myfrac#1#2{\left(\frac{#1}{#2}\right)}
\def\comment#1{\relax}

\begin{document}
\title{Do we see accreting magnetars in X-ray pulsars?}
%
%

\author{K.A. Postnov\inst{1}\fnsep\thanks{\email{pk@sai.msu.ru}} \and
        N.I. Shakura\inst{1}
\and
        A.Yu. Kochetkova\inst{1}
\and
	L. Hjalmarsdotter\inst{1}
}

\institute{Moscow M.V. Lomonosov State University, 
Sternberg Astronomical Institute, 13, Universitetskij pr., 119992 Moscow, Russia}

\abstract{%
Strong magnetic field of accreting neutron stars ($10^{14}$~G) is hard to probe by X-ray spectroscopy but can be indirectly inferred from spin-up/spin-down measurement in X-ray pulsars. The existing observations of slowly rotating X-ray pulsars are discussed. It is shown that magnetic fields of neutron stars derived from these observations (or lower limits in some cases) fall within the standard $10^{12}$-$10^{13}$~G range. Claims about the evidence for accreting magnetars are critically discussed in the light of recent progress in understanding of accretion onto slowly rotating neutron stars in the subsonic regime.
}
\maketitle
\section{Introduction}
\label{intro}
In the last decades, a class of isolated neutron stars (NSs) with high magnetic field $\sim 10^{14}-10^{15}$~G(magnetars) including soft gamma-ray
repeaters (SGRs) and anomalous X-ray pulsars (AXPs) is rapidly growing 
(see \cite{2008A&ARv..15..225M} for a review)\footnote{See online catalog
$http://www.physics.mcgill.ca/~pulsar/magne
tar/main.html$)
}. These sources exhibit 
slow NS spin periods ($P^*\sim 2-12$~s) and rapid spin-down rates ($\dot P\sim 5\times (10^{-10}-10^{-13}$~s/s), implying huge magnetic fields under the usual assumption of NS braking torque
describing by 
magneto-dipole losses $B\sim 10^{12}\sqrt{(P/1 \hbox{s})(\dot P/10^{-15}\hbox{s/s})}$~G. 
The high NS magnetic field can explain the observed properties of SGRs and AXPs in the frame of the magnetar model \cite{1995MNRAS.275..255T}. 
An alternative model, which can satisfactorily explain observations of SGRs and AXPs, 
assumes accretion onto a NS with the standard magnetic field 
from a supernova fall-back disk, see e.g. \cite{2013ApJ...764...49T}, \cite{2013IAUS..290...93A}
and references therein.

All known magnetars are single and young NSs, and their high magnetic fields should
reflect the initial magnetic field distribution of NSs. Initial field
as inferred from observations of radio pulsars by the population synthesis method 
using various model assumptions 
(see e.g. \cite{2006ApJ...643..332F}, \cite{2010MNRAS.401.2675P}) turn out to be 
broadly Gaussian-like distributed around the typical a few $10^{12}$~G value. 
The similar range $10^{12}-10^{13}$~G for magnetic fields of accreting NSs
has been found from X-ray spectroscopy of cyclotron resonance 
scattering features in X-ray pulsars \cite{2012MmSAI..83..230C}. 
Secular evolution of the NS magnetic field remains uncertain (e.g. 
\cite{2013arXiv1306.2156V}), but there seem to be no reasons to prohibit the 
birth of a strongly magnetized NS in a close binary system. Soon after the birth the NS may
become an accreting NS in a high-mass X-ray binary system. 

For accreting magnetized NSs, one can use, under certain assumptions, the 
plasma magnetospheric interaction to estimate the NS magnetic field.  
The accreting plasma brings angular momentum and exerts torques to the NS magnetosphere,
and there is magnetosphere-plasma coupling when plasma enters the magnetosphere,
therefore the accreting NS should spin-up or spin-down depending on 
the sign of the net torques applied to the magnetosphere, as observed \cite{1997ApJS..113..367B}. 
The torques $K_{su}, K_{sd}$ applied to the accreting NS depend on the NS magnetic field (dipole magnetic moment 
$\mu$), mass accretion rate $\dot M$ (as inferred from the observed X-ray luminosity $L_x\approx 
0.1 \dot Mc^2$), NS spin period $P^*$ (or angular frequency $\omega^*=2\pi/P^*$) and other
parameters depending on the regime of accretion. Then from the angular momentum conservation 
written as $I\dot\omega^*=K_{su}(\mu, \dot M...)-K_{su}(\mu, \dot M...)$ it is possible to 
estimate the NS magnetic field by assuming a specific form of the applied torques.

\section{Spin-up/spin-down torques in different accretion regimes}
\label{sec-1}

In different regimes of accretion, magnetospheric torques take different form 
(see Table \ref{t:torques}). Disk accretion takes place when the specific
angular momentum of matter near the NS magnetosphere exceeds the Keplerian 
one: $j_m(R_A)=\omega_m(R_A)R_A^2>j_K(R_A)=\sqrt{GMR_A}$; this is the usual case
hwen the optical star fills its Roche lobe. In the case of accretion from stellar wind, 
the specific angular momentum of gravitationally captured matter is $j_w=\eta\omega_B R_B^2$, 
where $\omega_B=2\pi/P_B$ is the binary orbital angular frequency, $P_B$ is the binary orbital period, $R_B=2GM/(v_w^2+v_{orb}^2)$ is the gravitational capture (Bondi) radius, 
$\eta$ is the numerical coefficient of order one, which can be negative 
or positive, so that on average, on the long range, the total angular momentum 
of captured matter can be close to zero \cite{1989MNRAS.238.1447H}. This 
intrinsic uncertainty 
strongly complicates the wind accretion case, and the conditions 
of the accretion disk formation  
in the wind-fed X-ray pulsars should be checked in each particular case using 
different observational data.
  
\begin{table*} 
\begin{tabular}{lccc}
\hline
Accretion regime & Spin-up torque, $K_{su}$ & Spin-down torque, $K_{sd}$ & Alfv\'en radius, $R_A$\\
\hline
\hline
Disk  & $\dot M\sqrt{GMR_A}$ & $\mu^2/R_c^3$ & $(\alpha\gamma\mu^2/\dot M)^{2/7}$\\
$R_A<R_c$\\
\hline
Quasi-spherical, & $\eta\dot M\omega_B R_B^2$ & $\eta\dot M\omega_B R_B^2$ & $(\mu^2/\dot M)^{2/7}$ \\
supersonic (Bondi)& $(\eta>0)$ & $(\eta<0)$\\
$L_x>4\times 10^{36}$~erg/s\\
\hline
Quasi-spherical, & $Z\dot M\omega_B R_B^2$ & $Z(1-z/Z)\dot M\omega^* R_A^2$ &  $(\mu^3/\dot M)^{2/11}$  \\ 
subsonic (settling)\\
$L_x<4\times 10^{36}$~erg/s\\
\hline
$\quad$Strong coupling: &$\mu^2/R_A^3$&$\mu^2/R_A^3$ \\
$\quad$$(B_p\sim B_\phi)$\\
$\quad$Moderate coupling:&$\frac{\mu^2}{R_A^3}\frac{(\omega_m-\omega^*)}{\omega_K(R_A)}$& $\frac{\mu^2}{\sqrt{{R_A^3}{R_c^3}}}, \quad \omega_m\ll \omega^*$\\
$\quad$(convection)\\
\hline
\end{tabular}
\label{t:torques}
\end{table*}

For wind-accreting X-ray pulsars with moderate and weak X-ray luminosity $L_x<4\times 10^{26}$~erg/s
the settling regime of subsonic quasi-spherical accretion sets in \cite{2012MNRAS.420..216S}, 
\cite{2013arXiv1302.0500S} (see also \cite{NSHmagbound}). For equilibrium pulsars
(in which on average $\dot \omega^*=0$), the estimation of
magnetic fields from spin-up/spin-down observations in this regime strongly depends on
stellar wind velocity, $\mu_{eq}\sim v_w^{-4}$, which is usually poorly known. 
 Equilibrium pulsars in the settling accretion regime have been discussed in \cite{NSHmagbound}.  
However, for 
non-equilibrium pulsars (e. g. those which demonstrate long-term spin-down) it is 
possible to obtain a lower limit of the NS magnetic field.

\section{Non-equilibrium X-ray pulsars}

The NS spin-up/spin-down equation in the settling accretion regime of quasi-spherical wind accretion (at $L_x<4\times 10^{36}$~erg/s) reads :
\beq{sd_eq1}
I\dot \omega^*= Z\dot M \omega_B R_B^2-Z(1-z/Z)\dot M R_A^2\omega^*\,,
\eeq 
(see \cite{NSHmagbound} for definitions of
the coefficients and derivation). 
The critical mass accretion rate at which $\dot \omega^*=0$ is 
\beq{dotmeq}
\dot M_{16,eq}\approx 478 (1-z/Z)^{11/4}\zeta \mu_{30}^3\myfrac{v_8}{\sqrt{\delta}}^{11}
\myfrac{P_b/10 \hbox{d}}{P_*/100 \hbox{s}}^\frac{11}{4}.
\eeq
At $\dot M<\dot M_{eq}$ the pulsar spins down, $\dot \omega^*<0$.  
From the simple fact that the spin down is stable, 
we may obtain a lower limit on the magnetic field in the case of quasi-spherical accretion with $\dot\omega^*<0$:
\beq{e:mulim}
\mu_{30}>\mu_{30, min}\approx 0.13
(1-z/Z)^{-\frac{11}{12}} \zeta^{-1/3}
\myfrac{\sqrt{\delta}}{v_8}^\frac{11}{3} \dot M_{16}^{1/3}
\myfrac{P_*/100 \hbox{s}}{P_b/10 \hbox{d}}^\frac{11}{12}\,.
\eeq 
It can be shown (see \cite{2013arXiv1302.0500S} for more detail) that at the spin-down
stage $\dot\omega^*$ reaches minimum at $\dot M_{cr}=\dot M_{eq}\times (3/7)^{11/4}\simeq 
0.1 \dot M_{eq}$:
\beq{e:omegadotsdmax}
\dot\omega^*_{sd,min}\approx -1.12\times 10^{-12}[\hbox{rad/s}^2] (1-z/Z)^{7/4} 
\tilde K K_1 K_3
\mu_{30}^{2}\myfrac{v_8}{\sqrt{\delta}}^{3}\myfrac{P^*}{100\hbox{s}}^{-7/4}
\myfrac{P_b}{10\hbox{d}}^{3/4}\,.
\eeq
Then, from the condition $|\dot \omega^*_{sd}|\le |\dot\omega^*_{sd,min}|$ 
follows a more interesting lower limit on the neutron star magnetic field:
\beq{e:mulim1}
\mu_{30}>\mu_{30, min}'\approx 0.94
\left|\frac{\dot\omega^*_{sd}}{10^{-12}\hbox{rad/s}^2}\right|^{1/2}(1-z/Z)^{-7/8} 
(\tilde K K_1 K_3)^{-1/2}
\myfrac{v_8}{\sqrt{\delta}}^{-3/2}\myfrac{P^*}{100\hbox{s}}^{7/8}
\myfrac{P_b}{10\hbox{d}}^{-3/8}.
\eeq
Note the weaker dependence of this estimate on the stellar wind velocity as compared to the inequality \Eq{e:mulim}.
If the accelerating torque can be neglected compared to the breaking torque (corresponding to the low X-ray luminosity limit $\dot M\ll \dot M_{eq}$), we find directly from \Eq{sd_eq1} that for accreting pulsars at spin down,  
\beq{e:sdonly}
\dot \omega^*_{sd}\approx - 0.75\times 10^{-12}[\hbox{rad/s}^2](1-z/Z)\tilde KK_1K_3\zeta^{-3/11} 
\mu_{30}^{13/11}\dot M_{16}^{3/11}\myfrac{P^*}{100\hbox{s}}^{-1}.
\eeq
From this we obtain a lower limit on the neutron star magnetic field that does not depend on the parameters of the stellar wind nor the binary orbital period:
\beq{e:mulim2}
\mu_{30}>\mu_{30, min}''\approx 1.27
\left|\frac{\dot\omega^*_{sd}}{10^{-12}\hbox{rad/s}^2}\right|^{11/13} 
(1-z/Z)^{-11/13}(\tilde KK_1K_3)^{-11/13}\zeta^{3/13}\dot M_{16}^{-3/13}
\myfrac{P^*}{100\hbox{s}}^{11/13}\,.
\eeq 

\section{Several examples}

Here we analyse the steady spin-down behavior in several slowly rotating moderate-luminosity  
X-ray pulsars (GX 1+4, SXP 1062, 4U 2206+54) within the framework of 
quasi-spherical settling accretion theory. The results are summarized in Table 2. 

\subsection{GX 1+4}

GX 1+4 has a spin period of  $P^*\approx 140$~s  and the donor is an MIII giant \cite{Davidsen_ea77}. 
The system has an orbital period of 1161 days \cite{Hinkle_ea06}. The donor is far 
from filling its Roche lobe and accretion onto 
the neutron star is by capture of the stellar wind of the companion. 
The system has a very interesting spin history. During the 1970's it was spinning up at the fastest rate ($\dot \omega_{su} \sim 3.8\cdot 10^{-11}$ rad/s) among the known X-ray pulsars at the time (e.g. \cite{Nagase89})). After several years of non-detections in the early 1980's, it reappeared again, now spinning down at a rate similar in magnitude to that of the previous spin-up. At present the source is steadily spinning down with an average spin down rate of 
$\dot omega^*_{sd}\approx -2.34\times 10^{-11}$~rad/s.  
A detailed spin-down history of the source is discussed in the recent paper \cite{2012A&A...537A..66G}. 
Using our model this behavior, as well as the observed inverse correlation 
between the instant torque applied to NS $-\dot\omega^*\sim L_x^{0.48}$ \cite{1997ApJ...481L.101C}, 
can be readily explained 
in the framework of quasi-spherical subsonic accretion (see also discussion of
GX 1+4 in \cite{2012MNRAS.420..216S},
\cite{2013arXiv1302.0500S}).

Clearly, GX 1+4 is not in equilibrium, 
so to derive a lower limit on the neutron star magnetic 
field from the observed value of $\dot \omega_{sd}$ we can use formulae from the previous Section. 
To avoid dependence on the uncertain stellar wind velocity, 
we shell neglect the spin-up torque, and from \Eq{e:mulim2} 
we get $\mu_{30,min}''\approx 40 (\tilde KK_1K_3)^{-11/13}\zeta^{3/13}$. 
Then, by assuming similarity of coupling parameters in all X-ray pulsars, it is safely to 
set the dimensionless factor $\tilde KK_1K_3\sim 10$ (as suggested by the analysis of 
equilibrium pulsars Vela X-1 and GX 301-2 in \cite{NSHmagbound}), thence $\mu_{30,min}''\approx 2.4$.
Somewhat higher value would be obtained from \Eq{e:mulim1}, under the same assumption of the
coupling coefficients: $\mu_{30,min}'\approx 9 (v_w/200\hbox{km/s})^{-3/2}$. Obviously, 
careful measurements of stellar wind velocity from th optical star 
and independent estimates of 
the NS magnetic field are highly needed for this source. 

\subsection{SXP 1062}

This recently discovered young X-ray pulsar in Be/X-ray binary system, located in a supernova remnant in the Small Magellanic Cloud. Its rotational period is $P^*\approx 1062$~s and it has a low X-ray luminosity of $L_x\approx 6\times 10^{35}$~erg/s \cite{2012MNRAS.420L..13H}. 
The source shows a remarkably high spin-down rate of $\dot \omega^*\approx -  1.6\times 10^{-11}$~(rad/s$^2$). 
Its origin is widely discussed in the literature (see e.g. \cite{2012A&A...537L...1H}, 
\cite{2012MNRAS.421L.127P}) and a possibly anormously high magnetic field of the neutron star has been suggested \cite{2012ApJ...757..171F}. In the framework of our model we use more conservative limits. 
Neglecting the spin-up torque \Eq{e:mulim2}, and assuming $\tilde KK_1K_3\sim 10$
for dimensionless coupling parameters \cite{NSHmagbound}, we get 
the magnetic field estimate $\mu_{30}>\mu_{30,min}''\approx 10$, which is independent on
the unknown stellar wind velocity and orbital parameters of the system.
This shows that the observed spin down can be explained by a magnetic field 
of the order of $10^{13}$~G, and thus we believe 
it is premature to conclude that the source is an accreting magnetar.

\subsection{4U 2206+54}

This slowly rotating pulsar has a period of $P^*=5560$~s and shows a spin-down rate of $\dot \omega_{sd}\approx -9.4 \times 10^{-14}$~rad/s \cite{2012MNRAS.425..595R}. The orbital period of the binary system is $P_b\simeq 19$~days \cite{2012MNRAS.425..595R}, and the measured stellar wind velocity is $v_W\approx 350$~km/s, abnormally low for an O9.5V \cite{2006A&A...449..687R} optical counterpart. 
The X-ray luminosity of the source is on average $L_x\simeq 2\times 10^{35}$~erg/s. 
A feature in the X-ray spectrum sometimes observed around 30 keV can be 
interpreted as a cyclotron line \cite{2004A&A...423..301T}, \cite{2004A&A...423..311M}, 
\cite{2006A&A...446.1095B}, \cite{2009MNRAS.398.1428W}. 
That gives an estimate of the magnetic field of the order of 
$B\sim (30/11.6)\times 1.3\approx  3.4\times 10^{12}$~G (taking into account the gravitational redshift close to the surface $1+z\sim 1.3$), and thus $\mu_{30}\approx 1.7$. Using this value of the magnetic field and neglecting the accelerating torque, from the formula in \Eq{e:sdonly} we obtain a lower limit on the parameter $\tilde KK_1K_3\gtrsim 20$, 
which is close to the coupling parameter values for the equilibrium pulsars Vela X-1 and GX 301-2
\cite{NSHmagbound}.  
If we consider the magnetic field to be unknown (see discussion in \cite{Reig_ea12}), and apply the formula\Eq{e:mulim2}, like in the case of GX 1+4, assuming moderate coupling with $\tilde KK_1K_3\sim 20$, 
we get the limit $\mu_{30}>\mu_{30,min}''\approx 0.6$, which is in agreement with standard neutron star magnetic field values. Note that applying our formulas for equilibrium pulsars would here give a magnetar value for the NS magnetic field \cite{2012MNRAS.425..595R}.
 
\begin{table*}
\label{T2}
 \centering
 \caption{} 
 $$
\begin{array}{lccc}
\hline
\hbox{Pulsars }&\multicolumn{3}{c}{\hbox{Non-equilibrium pulsars }}\\
\hline
& {\rm GX 1+4} &{\rm SXP1062}&{\rm 4U 2206+54}\\
\hline
\multicolumn{4}{c}{\hbox{Measured parameters}}\\
\hline
P^*{\hbox{(s)}} & 140 & 1062 &5560\\
P_B {\hbox{(d)}} & 1161 & \sim 300^\dag& 19(?)\\
v_{w} {\hbox{(km/s)}} & 200 & \sim 300^\ddag& 350\\
\mu_{30}&  ? & ? & 1.7\\
\dot M_{16} & 1 & 0.6 & 0.2\\ 
 \dot\omega^*_{sd} &  - 2.34 \cdot 10^{-11} & - 1.63 \cdot 10^{-11} & -9.4 \cdot 10^{-14}\\
\hline
\multicolumn{4}{c}{\hbox{Derived parameters}}\\
\hline
\tilde K K_1K_3\zeta^{-3/11} & & & \gtrsim 20\\
\mu_{30,min}''&\approx 2.4&\approx 10& \approx 0.6 \\
\hline
\end{array}
$$
$^\dag$ Estimate of the source's position in the Corbet diagram
$^\ddag$ Estimate of typical wind velocity binary pulsars containing Be-stars.
\end{table*}

\section{Conclusion}

Using theory of quasi-spherical subsonic accretion onto slowly rotating 
magnetized neutron stars \cite{2012MNRAS.420..216S}, 
we have obtain lower limits on the NS magnetic field in 
X-ray pulsars showng a long-term spin-down 
(GX 1+4, SXP 1062, 4U 2206+54). These limits in all cases 
turned out to be consistent with the standard value of NS magnetic fields $10^{12}-10^{13}$~G
as derived from analysis of radio pulsars \cite{2006ApJ...643..332F}, 
\cite{2010MNRAS.401.2675P} and observations of cyclotron resonance 
scattering features in spectra of X-ray pulsars \cite{2012MmSAI..83..230C}.
We conclude that present observations of spin-up/spin-down of X-ray pulsars
do not provide firm evidence for accreting magnetars, including the cases
of very slowly rotating low-luminosity pulsars (contrary to conclusions
in \cite{2012ApJ...757..171F}, \cite{2012MNRAS.425..595R}). 
The obtained lower limits of NS magnetic fields in 
X-ray pulsars do not exclude
high magnetic fields of accreting neutron stars, so we stress the 
need for further accurate timing and spectral observations of accreting X-ray pulsars.    

\textbf{Acknowledgements}. The authors acknowledge support from RFBR grant 12-02-00186a. 

%
\bibliography{wind}
%
%
%
%
%

\end{document}